\documentstyle[aps,twocolumn,prl,floats,epsf]{revtex}
\def\geqap{\,\raise 2pt \hbox{$>\kern-11pt \lower 5pt \hbox{$\sim$}$}\,}
\def\leqap{\,\raise 2pt \hbox{$<\kern-10pt \lower 5pt \hbox{$\sim$}$}\,}
\begin{document}
\draft
\twocolumn[\hsize\textwidth\columnwidth\hsize\csname @twocolumnfalse\endcsname
\title{Orbital Structure and Magnetic Ordering in Layered Manganites: \\ Universal Correlation 
and Its Mechanism}
\author{S.~Okamoto, S.~Ishihara, and S.~Maekawa}
\address{Institute for Materials Research, 
Tohoku University, Sendai 980-8577, Japan}
\date{\today}
\maketitle
\begin{abstract}
Correlation between orbital structure and magnetic ordering in bilayered manganites is examined. 
A level separation between the $3d_{3z^2-r^2}$ and $3d_{x^2-y^2}$ orbitals in a Mn ion 
is calculated in the ionic model for a large number of the compounds. 
It is found that the relative stability of the orbitals 
dominates the magnetic transition temperatures as well as 
the magnetic structures. 
A mechanism of the correlation between orbital and magnetism is investigated based on the theoretical 
model with the two $e_g$ orbitals under strong electron correlation. 
\end{abstract}
]
\narrowtext
%
%
Since the discovery of the colossal magnetoresistance (CMR), 
studies of manganites with cubic perovskite structure 
have been renewed theoretically and experimentally. 
Competition and cooperation 
between spin, charge and orbital degrees of freedom 
as well as lattice cause the dramatic 
changes of transport and magnetic properties.
Manganites with bilayered structure $A_{2-2x}B_{1+2x}$Mn$_2$O$_7$,  
where $A$ and $B$ are trivalent and divalent cations, respectively,  
are another class of CMR materials \cite{moritomo,kimura2}. 
Since an extremely large MR is observed near the transition 
from paramagnetic (PM) insulator to ferromagnetic (FM) metal, 
it has been considered that several concepts proposed in the cubic compounds  
are applicable to the bilayered ones. 
\par
In cubic manganites,  
one of the key factors dominating the magnetic orderings 
is the tolerance factor \cite{hwang}; 
a bending of a Mn-O-Mn bond  
decreases the hopping integral of carriers.
As a result, 
the ferromagnetic transition temperature $T_c$ decreases 
in the double exchange (DE) scenario. 
However, in bilayered manganites, 
the Mn-O-Mn bond angle is 
almost unchanged with changing cations  
and carrier concentration, as shown later, 
in spite of a wide variety of the magnetic structures.   
Various key factors dominating the magnetic ordering, which 
are not included in the DE model, were experimentally suggested, 
e.g. the antiferromagnetic (AFM) superexchange (SE) interaction \cite{perring},  
the local lattice distortion \cite{louca,dessau,medarde,doloc},  
the charge and orbital degrees of freedom and their orderings 
\cite{moritomo2,kimura} and so on. 
However, systematics in their correlations for a variety of compounds 
and their mechanisms still remain to be clarified. 
\par
In this letter,  we study the 
correlation between magnetic ordering  
and orbital structure in bilayered manganites. 
The two $e_g$ orbitals, i.e. the $3d_{3z^2-r^2}$ and $3d_{x^2-y^2}$ orbitals  
in a Mn$^{3+}$ ion 
split in the crystalline field of the bilayered structure 
and one of them is occupied by an electron. 
It is known that the occupied orbital controls the anisotropy of the magnetic interaction 
as well as its strength. 
The level separation between  
the orbitals is calculated in the ionic model 
for a large number of the compounds.  
We find a universal correlation between the relative stability of the orbitals 
and the magnetic transition temperatures as well as the magnetic structures. 
A mechanism of the correlation is investigated based on the theoretical model 
with the $e_g$ orbitals under strong electron correlation. 
\par
\begin{figure}
\epsfxsize=0.9\columnwidth
\centerline{\epsffile{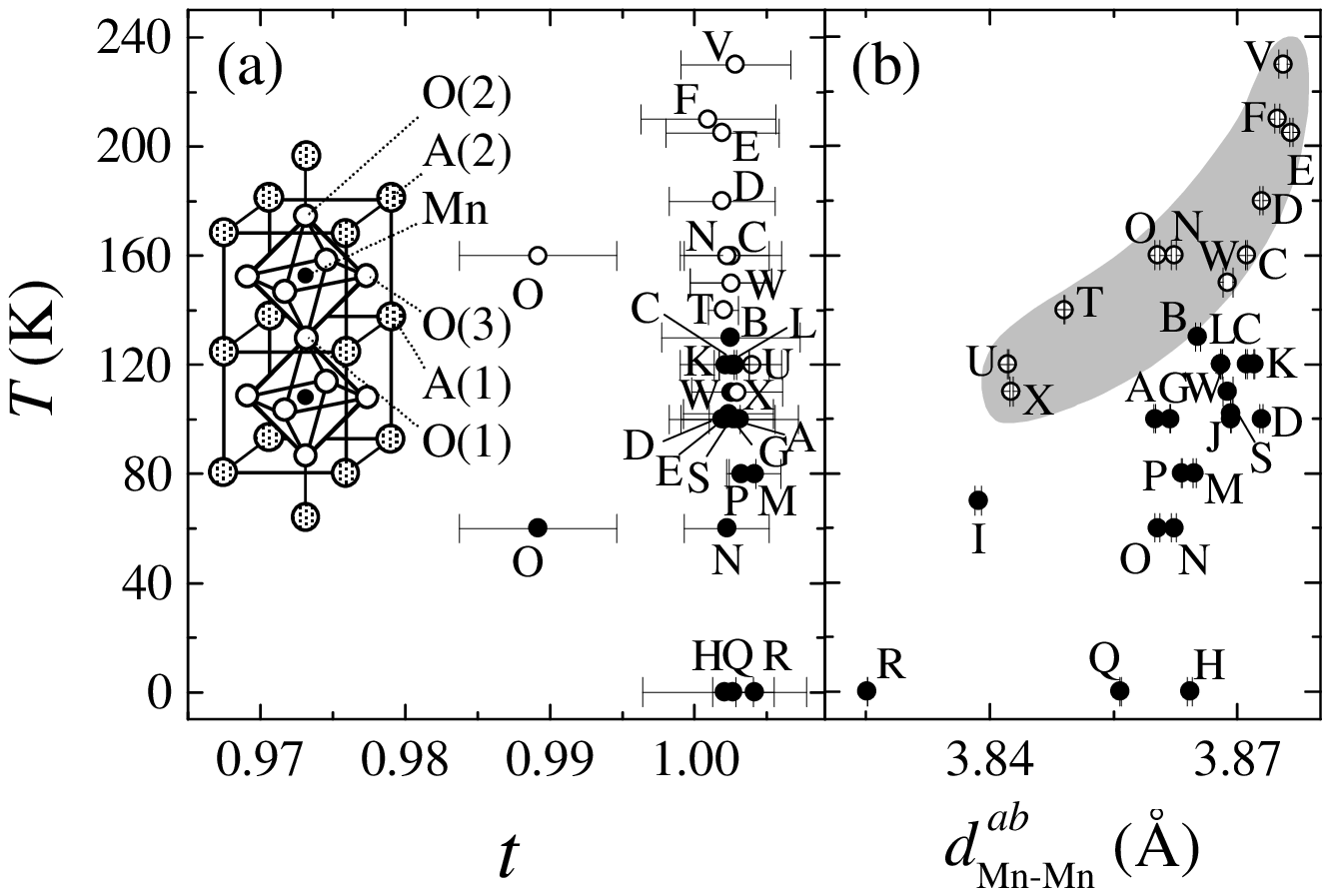}}
\caption{
$T_C$ and $T_N$ as functions of (a) $t$ 
and (b) $d_{\rm Mn-Mn}^{ab}$. 
Filled and open circles indicate $T_C$ and $T_N$, respectively. 
$t$ and $d_{\rm Mn-Mn}^{ab}$ are obtained from the structural data in 
the following compounds: 
A: La$_{1.4}$Sr$_{1.6}$Mn$_2$O$_7$                    (s)    \protect\cite{kubota}, 
B: La$_{1.3}$Sr$_{1.7}$Mn$_2$O$_7$                    (p)    \protect\cite{kubota},
C: La$_{1.2}$Sr$_{1.8}$Mn$_2$O$_7$                    (s)    \protect\cite{kubota},
D: La$_{1.1}$Sr$_{1.9}$Mn$_2$O$_7$                    (s)    \protect\cite{kubota},
E: La$_{1.04}$Sr$_{1.96}$Mn$_2$O$_7$                  (s)    \protect\cite{kubota},
F: LaSr$_2$Mn$_2$O$_7$                                (s)    \protect\cite{kubota}, 
G: La$_{1.4}$Sr$_{1.6}$Mn$_2$O$_7$                    (p)    \protect\cite{argyriou99}, 
H: Pr$_{1.4}$Ca$_{1.3}$Ba$_{0.3}$Mn$_2$O$_7$          (p)    \protect\cite{laffez}, 
I: Nd$_{1.4}$Ca$_{1.6}$Mn$_2$O$_7$                    (p)    \protect\cite{laffez}, 
J: La$_{1.4}$Sr$_{1.6}$Mn$_2$O$_7$                    (p)    \protect\cite{chi}, 
K: La$_{1.2}$Sr$_{1.8}$Mn$_2$O$_7$                       (p) \protect\cite{argyriou97}, 
L: La$_{1.2}$Sr$_{1.8}$Mn$_2$O$_7$                       (s) \protect\cite{akimoto}, 
M: La$_{1.2}$(Sr$_{0.8}$ Ca$_{0.2}$)$_{1.8}$Mn$_2$O$_7$   (s) \protect\cite{akimoto}, 
N: La$_{1.2}$(Sr$_{0.7}$ Ca$_{0.3}$)$_{1.8}$Mn$_2$O$_7$   (s) \protect\cite{akimoto}, 
O: La$_{1.2}$(Sr$_{0.6}$ Ca$_{0.4}$)$_{1.8}$Mn$_2$O$_7$   (p) \protect\cite{akimoto}, 
P: (La$_{0.8}$Nd$_{0.2}$)$_{1.2}$Sr$_{1.8}$Mn$_2$O$_7$   (s) \protect\cite{akimoto}, 
Q: (La$_{0.6}$Nd$_{0.4}$)$_{1.2}$Sr$_{1.8}$Mn$_2$O$_7$   (s) \protect\cite{akimoto}, 
R: Sm$_{1.2}$Sr$_{1.8}$ Mn$_2$O$_7$                      (p) \protect\cite{battle98}, 
S: La$_{1.2}$Sr$_{1.4}$Ca$_{0.4}$Mn$_2$O$_7$             (p) \protect\cite{shen}, 
T: NdSr$_2$Mn$_2$O$_7$                                   (p) \protect\cite{battle96}, 
U: Nd$_{1.1}$Sr$_{1.9}$Mn$_2$O$_7$                       (p) \protect\cite{battle96}, 
V: LaSr$_2$Mn$_2$O$_7$                                   (p) \protect\cite{seshadri}, 
W: LaSr$_{1.6}$Ca$_{0.4}$Mn$_2$O$_7$                     (p) \protect\cite{akimoto2}, 
X: NdSr$_2$ Mn$_2$O$_7$                                  (s) \protect\cite{akimoto2} 
where (s) and (p) indicate the single and polycrystalline samples, 
respectively. 
The inset of (a) shows a schematic picture of the bilayered structure.}
\label{fig:fig1}
\end{figure}

We first show that 
neither the tolerance factor nor the bond length 
governs $T_{C}$ and the N$\rm \acute e$el temperature $T_N$ for the A-type AFM ordering \cite{aafm}.
The tolerance factor in the bilayered crystals is defined by  
$
t=(d_{\rm O(1)-A(1)}+d_{\rm O(2)-A(2)}) /  (2\sqrt{2} d_{\rm Mn-O(3)})
$ 
where $d_{\rm A-B}$ is a bond length between A and B ions.  
The position of each ion is shown in the inset of Fig.~\ref{fig:fig1}(a). 
Being based on the structural data obtained by the neutron and x-ray diffraction experiments
\cite{kubota,argyriou99,laffez,chi,argyriou97,akimoto,battle98,shen,battle96,seshadri,akimoto2,notice}, 
we evaluate $t$ and the bond length between nearest neighboring (NN) Mn ions in the $ab$ plane 
$d_{\rm Mn-Mn}^{ab}$ for a variety of compounds. 
$T_C$ and $T_ N$ are plotted as functions of $t$ and 
$d_{\rm Mn-Mn}^{ab}$   
in Figs.~1 (a) and (b), respectively. 
Almost all $t$'s are located in a narrow region 
where $T_C$'s and $T_N$'s are distributed randomly.
In addition, $T_C$ is not correlated with $d_{\rm Mn-Mn}^{ab}$, either. 
Although $T_N$ increases with increasing $d_{\rm Mn-Mn}^{ab}$, 
this correlation is opposite to that predicted by the DE scenario where, 
with increasing the bond length, 
the hopping integral decreases and 
the FM interaction in the $ab$ plane decreases. 
We conclude that the DE model, which includes the change of the hopping integral 
caused by the change of 
the bond angle/length, can not explain $T_{C(N)}$.  
We also examined correlations between $T_{C(N)}$ and 
a number of other quantities: 
the torelance factor evaluated by the ionic radius, 
a Mn-O(3)-Mn bond angle, 
a Mn-O(1)-Mn bond length, Mn-O bond lengths, 
a lattice spacing between NN bilayers, 
lattice constants and the valence-bond sum for a Mn ion. 
However, there are not clear correlations between these parameters 
and $T_{C(N)}$.  
\par
Let us focus on the correlation between $T_{C(N)}$ and 
a relative stability of the $e_g$ orbitals. 
We employ the ionic model to examine the electronic energy-level structures. 
This model may be justified by 
the following considerations \cite{ohta}: 
(1) the manganites at $x=0$ are classified as charge-transfer type insulators
in which the ionic model provides a good starting point \cite{arima}. 
(2) The ionic property is predominant between bilayers.  
(3) The energy-level structure given by band-structure calculations
shows the same tendency with those by the ionic model \cite{dessau,band}. 
The energy levels of the $e_g$ orbitals split due to the electrostatic potential 
and one of the orbitals is occupied by an electron in a Mn$^{3+}$ ion. 
%

By using a large number of the structural 
data \cite{kubota,argyriou99,laffez,chi,argyriou97,akimoto,battle98,shen,battle96,seshadri,akimoto2,notice}, 
we calculate the Madelung potential for a hole in the $3d_{3z^2-r^2}$ and $3d_{x^2-y^2}$ orbitals 
at site $j$ defined by 
\begin{equation}
V_{3z^2-r^2}={1 \over 2} \bigl \{V(\vec r_j+r_d \hat z)+V(\vec r_j-r_d \hat z) \bigr \}  , 
\end{equation}
and 
\begin{equation}
V_{x^2-y^2}=V(\vec r_j+r_d \hat x) , 
\end{equation} 
respectively \cite{ishihara}.  
Here, $V(\vec r_j)$ is given by 
\begin{equation}
V(\vec r_j)=\sum_{i \ne j} {Z_ie^2 \over |\vec r_j -\vec r_i| }, 
\label{eq:sum}
\end{equation}
with a point charge $Z_ie$ at site $i$ and the position $\vec r_i$ of the site.  
$r_d$(=0.42\AA) is the radius of a Mn $3d$ orbital 
where its radial charge density becomes maximum \cite{slater} 
and $\hat z (\hat x)$ is the unit vector in the $z(x)$ axis. 
The Ewald method is used for the lattice summation. 
$Z_i$'s for Mn and O ions and a cation at $A$ site are chosen to be $3+x$, $-2$ and $(8-2x)/3$, respectively. 
The difference of the potentials  
\begin{equation}
\Delta V= V_{3z^2-r^2}-V_{x^2-y^2} , 
\end{equation}
represents the relative stability of the orbitals; 
with increasing $\Delta V$,  
the energy level of the $3d_{3z^2-r^2}$ orbital for an electron relatively decreases. 
\par
\begin{figure}
\epsfxsize=0.8\columnwidth
\centerline{\epsffile{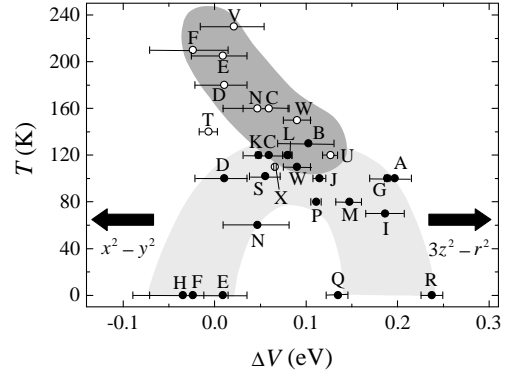}}
\caption{$T_C$ and $T_N$ as functions of  
$\Delta V$. 
Filled and open circles indicate $T_C$ and $T_N$, respectively.
$\Delta V$'s are calculated for the same compounds in Fig.~1. 
Note that in the region with large positive (negative) $\Delta V$, 
the $3d_{3z^2-r^2}$ ($3d_{x^2-y^2}$) orbital is occupied by an electron.}
\label{fig:fig2}
\end{figure}

$T_N$ and $T_ C$ 
are plotted as functions of $\Delta V$ in Fig.~\ref{fig:fig2} where  
the structural data at room temperature are used. 
Broad shades are drawn by considering experimental errors. 
It is clearly shown that both $T_C$ and $T_N$ are correlated with $\Delta V$; 
$T_N$ increases with decreasing $\Delta V$ and there is an optimal value of 
$\Delta V$($\sim 0.08$ eV) for $T_C$. 
We estimate the strength of the correlation between $T_N$ and $\Delta V$ 
by using the correlation coefficient:  
$r={1\over N} \sum_l$
$(T_{Nl}-\overline{T_N})(\Delta V_l-\overline{\Delta V}) / 
(\sigma_{T_N}\sigma_{\Delta V} )
$ 
where $l$ indicates a sample and $N$ is the number of samples. 
$\overline{\Delta V}$ ($\overline{T_N}$) and $\sigma_{\Delta V}$ ($\sigma_{T_N}$)
are the mean value and the standard deviation of $\Delta V$ ($T_N$), respectively.  
We obtain $r=-0.89 \pm 0.11$ for single crystal samples 
and $r=-0.15 \pm 0.04$ for all samples including polycrystals. 
One might think that the $T_{C(N)}$ v.s. $\Delta V$ curve in Fig.~\ref{fig:fig2} 
just reflects the relation between $T_{C(N)}$ and $x$ in 
La$_{2-2x}$Sr$_{1+2x}$Mn$_2$O$_7$ (LSMO)\cite{kubota,medarde}.  
However, when we pay attention to $T_{C(N)}$'s for samples with the same $x$ 
(e.g., the samples C and K-R), we notice that the correlation remains. 
The correlation between $T_N$ and $\Delta V$ explains that between $T_N$ and 
$d_{\rm Mn-Mn}^{ab}$ shown in Fig.~1(b), 
since $\Delta V$ is a decreasing function of $d_{\rm Mn-Mn}^{ab}$ in the 
region, 3.84${\rm \AA} < d^{ab}_{\rm Mn-Mn} <$3.88$\rm \AA$. 
%
%
\par
In Fig.~3, we present the magnetic phase diagram at $T=0$ as a function of 
$\Delta V$ and $x$. 
The structural data at room temperature are used.
Symbols connected by dotted lines correspond to a series of 
LSMO with $x=0.3-0.5$ \cite{kubota}. 
In addition, $\Delta V$'s calculated by using the data below $T_{C(N)}$ in LSMO are also plotted. 
Bold arrows indicate the change of the Madelung potential 
with changing temperature from $T>T_{C(N)}$ to $T<T_{C(N)}$: 
\begin{equation}
\delta (\Delta V) \equiv \Delta V(T<T_{C(N)})-\Delta V(T>T_{C(N)}) . 
\label{eq:deltv}
\end{equation}
We find that 
the magnetic structures are governed by $\Delta V$ and $x$; 
the FM (A-type AFM) phase is located in the region with 
smaller (larger) $x$ and moderate (smaller) $\Delta V$.  
Let us focus on $\delta (\Delta V)$ in LSMO. 
$\delta (\Delta V)$'s are negative at $x=0.3$ and 0.35. 
The absolute value of $\delta (\Delta V)$
gradually decreases with increasing $x$ and 
$\delta (\Delta V)$ becomes a small positive value at $x=0.4$.
Below $T_C$, $\Delta V$ seems to approach to the optimal value of $\Delta V \sim 0.08$ where 
$T_C$ becomes maximum as seen in Fig.~\ref{fig:fig2}. 
On the other hand, 
$\delta (\Delta V)$'s are negative at $x$=0.45 and 0.48 where  
the A-type AFM structure appears.
The orbital structure and its stability in the FM phase have been studied by 
measuring the striction in Ref.~\onlinecite{kimura}. 
The difference of the Mn-O bond lengths between PM and FM 
states was reported in Ref.~\onlinecite{medarde}. 
These experimental results are consistent with the present results of $\delta (\Delta V)$ in Fig.~3. 
%
%
%
\begin{figure}
\epsfxsize=0.8\columnwidth
\centerline{\epsffile{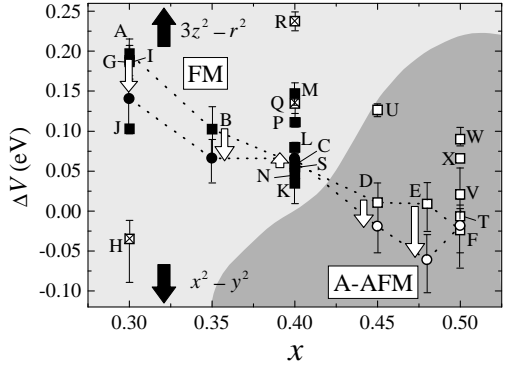}}
\caption{The magnetic phase diagram at $T=0$ as a function of $\Delta V$ and $x$.
Filled, open and crossed squares indicate the FM, A-type AFM and 
PM samples, respectively. 
$\Delta V$'s for filled and open circles are obtained  
from the data below $T_{C}$ and $T_N$, respectively.
Symbols connected by dotted lines indicate a series of LSMO. 
Bold arrows show $\delta (\Delta V)=\Delta V(T<T_{C(N)})-\Delta V(T>T_{C(N)})$. 
$\Delta V$'s are calculated for the same compounds in Fig.~1.
Note that in the region with large positive (negative) $\Delta V$, 
the $3d_{3z^2-r^2}$ ($3d_{x^2-y^2}$) orbital is occupied by an electron.}
\label{fig:fig3}
\end{figure}
\par
Now we theoretically investigate a mechanism of the correlation between 
magnetic ordering and orbital structure. 
We start with the following Hamiltonian \cite{ishihara,ishihara2,hamiltonian}:  
\begin{equation}
{\cal H}={\cal H}_t+{\cal H}_J+{\cal H}_H+{\cal H}_{AF}+{\cal H}_z .  
\label{eq:hamiltonian}
\end{equation}
Instead of the bilayered structure, 
the simple tetragonal lattice consisting of Mn ions is considered.  
In this model, the magnetic structure with FM and AFM 
alignments perpendicular and parallel to the $c$ axis, respectively, 
corresponds to the A-type AFM structure \cite{aafm}. 
In each Mn ion, the two $e_g$ orbitals are introduced and 
the $t_{2g}$ electrons are treated as a localized spin with $S=3/2$. 
The first two terms in Eq.~(\ref{eq:hamiltonian}) correspond to the so-called $t$- and $J$-terms 
in the $tJ$ model, respectively, with the  two $e_g$ orbitals under strong electron correlation.  
The third and fourth terms describe the Hund coupling between $e_g$ and $t_{2g}$ spins
and the AFM SE interaction between $t_{2g}$ spins, respectively. 
The splitting of the energy levels between $3d_{3z^2-r^2}$ and $3d_{x^2-y^2}$ 
orbitals is represented by the last term:  
$
{\cal H}_z=-\Delta \sum_i T_{iz} 
$ 
where the pseudospin operator is given by 
$\vec T_{i}={1 \over 2} \sum_{\gamma \gamma' \sigma} 
d_{i \gamma \sigma}^\dagger \vec \sigma_{\gamma \gamma'} d_{i \gamma' \sigma}$ 
with $d_{i \gamma \sigma}$ being the annihilation operator of an $e_g$ electron 
at site $i$ with spin $\sigma$ and orbital $\gamma$. 
The $+(-)$ eigenstate of $T_{iz}$ corresponds to the state where 
the $3d_{3z^2-r^2}$ ($3d_{x^2-y^2}$) orbital is occupied by an electron. 
The anisotropies of the hopping integral and the SE interactions 
due to the layered structure are considered. 
The explicit expression and derivation of the Hamiltonian are 
presented in Refs.~\cite{ishihara} and \cite{ishihara2}. 
\par
%
%
%
\begin{figure}
\epsfxsize=0.8\columnwidth
\centerline{\epsffile{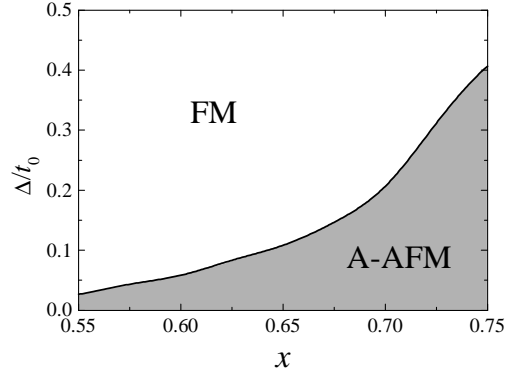}}
\caption{The calculated magnetic phase diagram at $T=0$ as a function of $\Delta$ and $x$. 
Note that in the region with large positive (negative) $\Delta$, 
the $3d_{3z^2-r^2}$ ($3d_{x^2-y^2}$) orbital is occupied by an electron.
}
\label{fig:fig4}
\end{figure}
The calculated magnetic phase diagram at $T=0$ is presented in Fig.~4 \cite{hamiltonian,maezono} 
where the mean field approximation is adopted. 
We note that the phase diagram derived in this approximation 
explains that in the cubic manganites \cite{okamoto}. 
The characteristic features shown in Fig.~3  
are well reproduced by the present theory; 
the A-type AFM phase appears in the region with higher $x$ 
and smaller $\Delta $ than that of the FM one. 
The range of the horizontal axis in Fig.~4 is larger than 
that in Fig.~3 by about $0.25$. 
This discrepancy may be attributed to the neglect of the orbital fluctuation \cite{ishihara3}.   
However, the characteristics of the phase diagram   
are insensitive to the parameters in the model. 
In the FM (A-type AFM) phase, the orbitals are uniformly aligned 
with $ 0<\theta <0.72\pi$ ($0.72\pi<\theta<\pi$)  
where $\theta$ describes the orbital state as 
$|\theta \rangle=$$\cos({\theta \over 2})|3d_{3z^2-r^2}\rangle$$-\sin({\theta \over 2})|3d_{x^2-y^2} \rangle$. 
The present results suggest that 
a dimensionality of the FM interaction is controlled by 
the orbital structure; 
in the A-type AFM phase, 
the FM ordering in the $ab$ plane 
is caused by the DE interaction, while the AFM in the $c$ direction is by the AFM SE. 
When the $3d_{x^2-y^2}$ orbital is stabilized,  
the DE interaction in the $ab$ plane ($c$ direction) becomes strong (weak)  
and the A-type AFM phase appears \cite{maezono1}. 
A mixing of the orbitals is essential in the FM phase 
where the FM interaction overcomes the AFM SE one in the three directions. 
We note that, in Fig.~4, the FM phase appears not around $\Delta=0$ 
but in a region of $\Delta>0$,  
since the anisotropy in the hopping integral due to the layered structure 
stabilizes the $3d_{x^2-y^2}$ orbital more than $3d_{3z^2-r^2}$. 
It is worth to mention the change of the orbital structure associated with 
the magnetic ordering: 
By utilizing the mean filed approximation at finite temperature, 
we compare the orbital structures above and below the magnetic 
transition temperatures. 
It is found that (1) 
there is an optimal mixing of the orbitals for the FM state 
and the orbital structure tends to approach to this structure below $T_C$ 
and 
(2) the $3d_{x^2-y^2}$ orbital structure is stabilized below $T_N$. 
The theoretical results are consistent with $\delta (\Delta V)$'s shown in Fig.~3 
by considering that the change of $\Delta V$ 
associated with the magnetic ordering is caused by that of the orbital structure. 
\par
In summary, 
we examine correlation between magnetic ordering and orbital structure 
in bilayered manganites. 
A relative stability of the $e_g$ orbitals is 
investigated by calculating 
the Madelung potentials   
in a large number of the bilayered compounds. 
We find that 
the A-type AFM structure and 
the $3d_{x^2-y^2}$ orbital one are stabilized cooperatively 
and there is 
an optimal mixing between the $3d_{3z^2-r^2}$ and $3d_{x^2-y^2}$ orbitals for the FM ordering. 
A theory with the two $e_g$ orbitals under strong electron correlation 
explains a mechanism of the universal correlation between orbital and magnetism. 
\par
The authors would like to thank Y.~Moritomo, T.~Akimoto, Y.~Tokura, 
T.~Kimura, Y.~Endoh, K.~Hirota, M.~Kubota and G.~Khaliullin for their valuable discussions. 
This work was supported by CREST, NEDO 
and Grant-in-Aid for Scientific Research Priority Area 
from the Ministry of Education, Science and Culture of Japan.  
S. O. acknowledges the financial support of JSPS.  
Part of the numerical calculation was performed in the HITACS-3800/380 
superconputing facilities in IMR, Tohoku Univ.

\begin{references}
\bibitem{aafm}
The following two layered AFM structures have been observed \protect\cite{kubota,argyriou99}: 
AFM-I consisting of the FM spin ordering in the plane and the AFM (FM) one along the $c$ axis 
within (between NN) bilayers, 
and 
AFM-II consisting of the FM in the plane and the FM (AFM) along the $c$ axis within (between NN) bilayers. 
%
Since the intrabilayer magnetic coupling is much larger than the interbilayer one \cite{fujioka}, 
we term the AFM-I structure the A-type AFM one and 
regard the AFM-II  as the FM.
\bibitem{notice}
For the samples G, J, R, T and U 
where the two chemically distinct phases are observed, 
the structural data of the major phase are adopted.
\bibitem{hamiltonian}
In the Hamiltonian, we consider the SE interactions $J_{1,2}^{ab(c)}$ between $e_g$ electrons, the AFM SE one 
$J_{AF}^{ab(c)}$ between $t_{2g}$ spins and 
the hopping integral $t_0^{ab(c)}$ of $e_g$ electrons 
where the subscript $ab$ $(c)$ indicates the interaction in 
the $ab$ plane (the $c$ direction) \protect\cite{ishihara,ishihara2}. 
These values are chosen as 
$J_{1(2)}^{ab}/t_0^{ab}=0.25$ (0.075), $J_{AF}^{ab}/t_0^{ab}=0.002$ and   
$t_0^{ab}/t_0^{c}=$$\sqrt{J_{1,2}^{ab}/J_{1,2}^{c}}=\sqrt{J_{AF}^{ab}/J_{AF}^{c}}=1.5$ 
in the numerical calculation. $t_0^{ab}$ is represented as $t_0$ for simplicity.
\end{references}
\end{document}